# Axial Thermal Transport in Transition Metal Chalcogenide Nanowires from Machine-Learning Molecular Dynamics Simulations

Wenwu Jiang[1,*] and Hekai Bu[2]

[1]*Department of Modern Mechanics, University of Science and Technology of China, Hefei, Anhui 230026, China*

[2]*College of Physical Science and Technology, Bohai University, Jinzhou, China*

[*]Corresponding author. Email: wwjiang96@ustc.edu.cn

**ABSTRACT**

Transition metal chalcogenide $M_6X_6$ nanowires (M = Mo or W; X = S, Se, or Te) are promising one-dimensional (1D) building blocks for nanowire bundles and arrays, but how elemental composition controls their intrinsic axial heat transport remains unclear. Here we develop a multicomponent neuroevolution potential (NEP) from first-principles reference data and use homogeneous nonequilibrium molecular dynamics (HNEMD) simulations to quantify the thermal conductivity of six $M_6X_6$ nanowires. The $Mo_6X_6$ series show a nonmonotonic composition dependence, whereas the thermal conductivity of the $W_6X_6$ series decreases monotonically with increasing chalcogen mass. Frequency-resolved thermal conductivity and phonon mean-free path analyses show that axial heat conduction is dominated by low-frequency phonons. Vibrational density of states and spectral energy density analyses further reveal that the contrasting trends arise from a competition between metal-chalcogen vibrational matching and phonon softening induced by heavier chalcogen atoms. A phenomenological model incorporating the combined atomic mass and the relative mass difference between the metal and chalcogen atoms reproduces these trends. These results link elemental composition, phonon characteristics and axial thermal conductivity in $M_6X_6$ nanowires, providing an atomistic basis for selecting heat-conducting units in nanowire bundles and arrays.

## 1. INTRODUCTION

One-dimensional (1D) nanowire bundles and arrays have attracted increasing interest as anisotropic building blocks for nanoscale electronics and flexible devices.[1-6] Their high aspect ratios, direction-dependent transport properties and structural flexibility make them attractive for integrated low-dimensional architectures. In such devices, however, localized Joule heating and inefficient heat dissipation can degrade performance and reliability.[7, 8] Understanding heat transport in nanowire-based assemblies is therefore essential for their practical implementation.[5-12] Because the thermal response of an assembly is set by both the intrinsic axial transport of individual nanowires and the interwire junctions or interfaces,[6, 13] identifying highly thermally conductive individual nanowires is the necessary first step toward thermally reliable nanowire architectures.

Transition metal chalcogenide $M_6X_6$ nanowires, where M = Mo or W and X = S, Se, or Te, provide a useful platform for isolating the intrinsic heat-conducting units of nanowire assemblies.[14-20] These materials possess 1D atomic structures, subnanometer diameters, and weak van der Waals (vdW) interactions between neighboring wires, making them suitable model systems as well as realistic components for assembled nanowire architectures.[17-19, 21] Experimentally, $Mo_6S_6$ nanowires have been shown to assemble into trimer bundles, whereas $Mo_6Te_6$ nanowires can form arrays and bundles stabilized by vdW interactions on different substrates.[22-24] In addition, chemical vapor deposition has enabled the wafer scale synthesis of transition metal telluride nanowires, demonstrating the feasibility of fabricating large area networks and bundles composed of individual atomic wires.[25, 26] Despite these advances in synthesis and assembly, the thermal transport properties of the $M_6X_6$ family remain largely unexplored. Previous molecular dynamics (MD) work on $Mo_6S_6$ mainly focused on the mechanical modulation of phonon thermal conductivity under tension and torsion, whereas how the elemental composition of $M_6X_6$ nanowires governs phonon transport across Mo/W and S/Se/Te combinations remains unclear. This gap limits the selection of efficient intrinsic heat-conducting units for $M_6X_6$ nanowire bundles and arrays.

Here we developed a machine-learned interatomic potential, specifically a neuroevolution potential (NEP),[27] and performed MD simulations to systematically investigate the intrinsic axial thermal transport of six $M_6X_6$ nanowires, including $Mo_6S_6$, $Mo_6Se_6$, $Mo_6Te_6$, $W_6S_6$, $W_6Se_6$, and $W_6Te_6$. The calculated thermal conductivities exhibit a pronounced dependence on elemental composition. Notably, the Mo-based and W-based nanowires follow distinct trends rather than a simple monotonic dependence on atomic mass. $Mo_6Se_6$ exhibits a higher thermal conductivity than both $Mo_6S_6$ and $Mo_6Te_6$, whereas the thermal conductivity of the $W_6X_6$ series decreases monotonically from sulfide to telluride. Spectral thermal conductivity, phonon spectral energy density (SED),[28] and vibrational density of states (DOS) analyses reveal that these contrasting trends arise from the competition between vibrational matching of the metal and chalcogen atoms and the phonon softening induced by heavier chalcogen atoms. A simple phenomenological model incorporating the combined atomic mass and the relative mass difference between the metal and chalcogen atoms further rationalizes the observed composition dependence.

## 2. MATERIALS AND METHODS

### 2.1. Development of the NEP Model

To describe the interatomic interactions in $M_6X_6$ nanowires, we developed a multi-component NEP[27] based on first-principles reference data. The NEP model includes five atomic species (Mo, W, S, Se, and Te), and was constructed for six representative $M_6X_6$ nanowires, including $Mo_6S_6$, $Mo_6Se_6$, $Mo_6Te_6$, $W_6S_6$, $W_6Se_6$, and $W_6Te_6$, as shown in **Figure 1**a. The reference configurations were generated by combining randomly perturbed structures and density functional theory molecular dynamics (DFT-MD) snapshots, thereby sampling both locally distorted and thermally accessible atomic environments. Further details of the dataset generation are provided in Section 1 of the Supporting Information. After constructing the initial training configurations, first-principles calculations for the reference dataset were performed using the Vienna Ab initio Simulation Package (VASP)[29, 30] within the projector augmented-wave (PAW)[31] framework and a plane-wave basis set. The exchange-correlation functional was described using the Perdew-Burke-Ernzerhof (PBE)[32] form of the generalized gradient approximation. The reference energies, atomic forces, and virial tensors used for NEP training were calculated with a plane-wave cutoff energy of 850 eV. The electronic self-consistent convergence criterion was set to $10^{-7}$eV, and the maximum number of electronic self-consistent steps was set to 150.

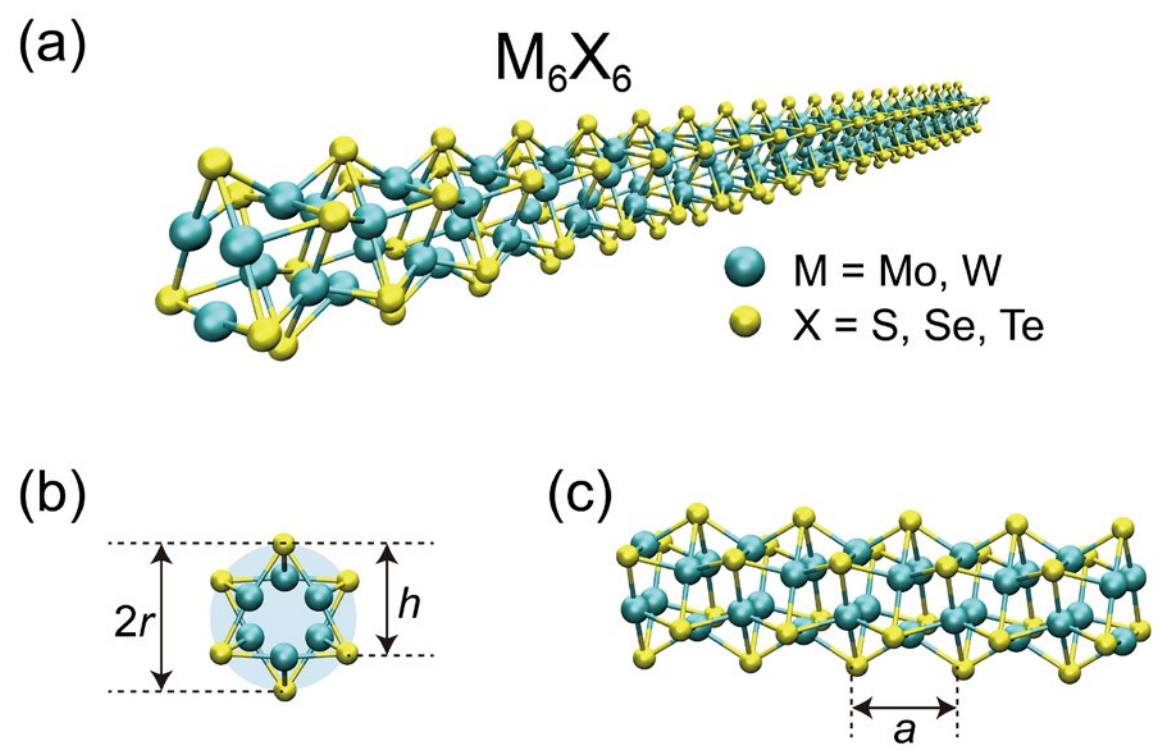

**Figure 1.** Atomic structure and geometric parameters of $M_6X_6$ nanowires. (a) Representative atomistic model of 1D $M_6X_6$ nanowires, where M denotes Mo or W and X denotes S, Se, or Te. The M and X atoms are rendered in cyan and yellow, respectively. (b) Cross-sectional view of the nanowire. Here, $2r$ denotes the diameter of the effective circular cross section used to define the effective cross-sectional area $A = \pi r^2$, and $h$ denotes the transverse structural height of the nanowire. (c) Side view along the nanowire, where $a$ is the axial lattice constant.

The neural network potential was constructed using the fourth-generation neuroevolution potential (NEP4)[33] framework. Within this formalism, the site energy $U_i$ of atom $i$ is expressed by a feedforward neural network with a single hidden layer containing $N_{\text{neu}}$ neurons:[33]

$$U_i = \sum_{\mu=1}^{N_{\text{neu}}} w_{\mu}^{(1)} \tanh\left(\sum_{\nu=1}^{N_{\text{des}}} w_{\mu\nu}^{(0)} q_{\nu}^{i} + b_{\mu}^{(0)}\right) + b^{(1)}, \quad (1)$$

where $q_{\nu}^{i}$ denotes the $N_{\text{des}}$-dimensional descriptor vector that characterizes the local atomic environment of atom $i$, and $\tanh(x)$ is the activation function. The quantities $w^{(0)}$, $w^{(1)}$, $b^{(0)}$, and $b^{(1)}$ are the trainable weights and biases of the neural network.

The local descriptor set includes both radial and angular contributions. Model parameters were determined by minimizing a loss function composed of the weighted root-mean-square errors (RMSEs) of energies, forces, and virials, together with a regularization term. To determine an appropriate descriptor cutoff radius, we systematically compared NEP models with the radial and angular cutoff radius of 3.0, 3.5, 4.0, and 4.5 Å, as shown in **Figure 2**. Increasing the cutoff radius from 3.0 to 3.5 Å substantially reduced the RMSEs of energy, force, and virial, indicating that a 3.0 Å cutoff is insufficient to fully capture the relevant local atomic environments in $M_6X_6$ nanowires. Further increasing the cutoff radius from 4.0 to 4.5 Å led only to marginal improvements, while increasing the computational cost. Therefore, the cutoff radii of 4.0 Å was adopted for both the radial and angular descriptors as a balance between accuracy and efficiency.

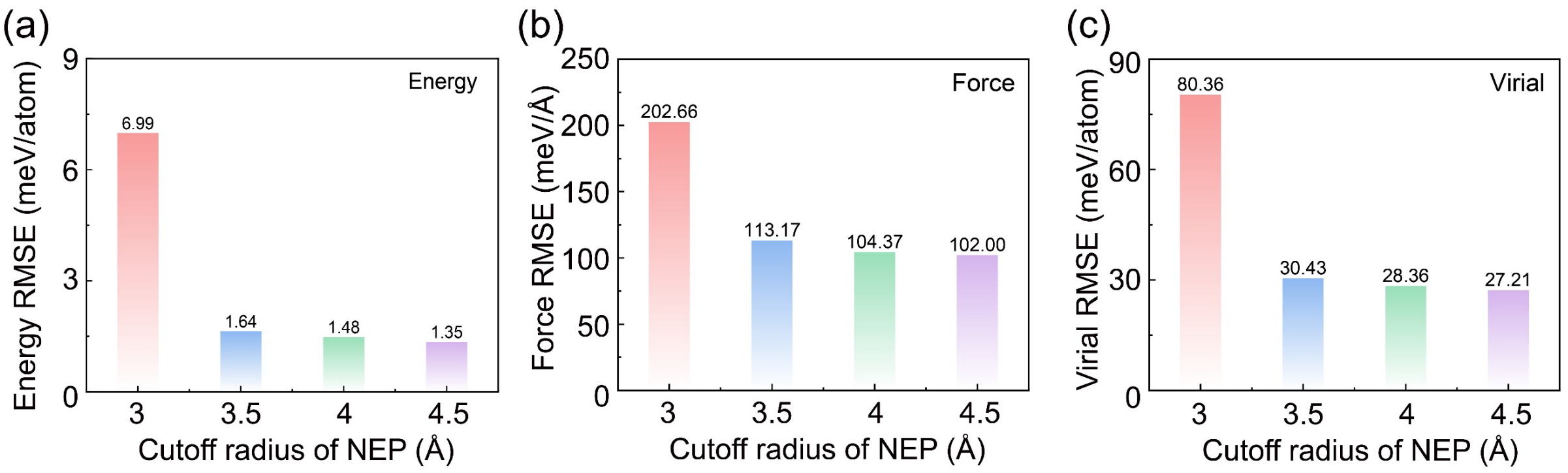

**Figure 2.** Effect of descriptor cutoff radius on the accuracy of the NEP model. RMSEs of the NEP predictions for (a) energy, (b) atomic force, and (c) virial tensor as a function of the descriptor cutoff radius. The red, blue, green, and purple bars correspond to cutoff radii of 3.0, 3.5, 4.0, and 4.5 Å, respectively.

For each descriptor type, nine radial basis functions were employed, with each basis function further expanded as a linear combination of 13 Chebyshev polynomials. Following the atomic cluster expansion strategy for the construction of angular descriptors, three-body and four-body correlations described by spherical harmonics were included up to angular orders $l = 4$ and $l = 2$, respectively. The final NEP model was trained for 500,000 generations using a full-batch optimization scheme.

## 2.2. HNEMD Method

The HNEMD method provides a mechanical analogue of thermal transport and, within the framework of linear response theory, enables an efficient evaluation of thermal transport coefficients.[34] It has been widely recognized as one of the most computationally efficient molecular dynamics approaches for characterizing diffusive heat transfer. In the HNEMD formalism, the system is driven out of equilibrium by an external field, and the corresponding external driving force acting on atom $i$ is written as:[34]

$$\boldsymbol{F}_i^{\mathrm{ext}} = E_i \boldsymbol{F}_e + \boldsymbol{F}_e \cdot \mathbf{W}_i, \tag{2}$$

where $\boldsymbol{F}_e$ is a vectorial driving force parameter with units of inverse length, $E_i$ is the total energy of atom $i$, and $\mathbf{W}_i$ is the virial tensor of atom $i$, defined as:

$$\mathbf{W}_i = \sum_{j \neq i} \left( \frac{\partial U_j}{\partial \boldsymbol{r}_{ji}} \otimes \boldsymbol{r}_{ij} \right). \tag{3}$$

Here, $U_i$ is the potential energy assigned to atom $i$, and $\boldsymbol{r}_{ij} \equiv \boldsymbol{r}_j - \boldsymbol{r}_i$, with $\mathbf{r}_i$ being the position vector of atom $i$. Under the action of the external field, the system reaches a nonequilibrium steady state, in which the ensemble average of the heat current $\langle \boldsymbol{J} \rangle$ satisfies

$$\frac{\langle J^\alpha \rangle}{TV} = \sum_\beta \kappa^{\alpha\beta} F_e^\beta, \tag{4}$$

where $T$ and $V$ denote the system temperature and volume, respectively, and $\kappa^{\alpha\beta}$ is the $\alpha\beta$ component of the thermal conductivity tensor. In the present work, we focus on axial thermal transport along the *z* direction of the 1D $M_6X_6$ nanowires and therefore evaluate the diagonal component $\kappa^{zz}$ as:

$$\kappa^{zz} = \frac{\langle J^z \rangle}{TVF_e^z}. \tag{5}$$

The magnitude of the driving field $F_e^z$ was determined by systematically testing a range of $F_e$ values (see Section 2 of the Supporting Information). The selected values ensure that the induced heat current remains within the linear response regime while providing a sufficiently clear signal. Specifically, $F_e^z$ was set to $7.5 \times 10^{-6}$ Å$^{-1}$ for $Mo_6S_6$ and $W_6S_6$, and to $5 \times 10^{-6}$ Å$^{-1}$ for the other four $M_6X_6$ nanowires.

An important advantage of the HNEMD method is that it allows a spectral decomposition of the thermal conductivity. Specifically, the in-plane thermal conductivity along the $z$ direction can be expressed as:

$$\kappa^{zz} = \int_0^\infty \frac{\mathrm{d}\omega}{2\pi} \kappa^{zz}(\omega), \tag{6}$$

where the frequency-resolved thermal conductivity is given by

$$\kappa^{zz}(\omega) = \frac{2}{TVF_{\mathrm{e}}^{z}}\int_{-\infty}^{\infty} \mathrm{d}t\, \mathrm{e}^{i\omega t} K^{z}(t). \tag{7}$$

Here, $K^{z}(t)$ is the $z$-component of the virial–velocity correlation function vector $\boldsymbol{K}(t)$, defined as:

$$\boldsymbol{K}(t) = \sum_i \langle \mathbf{W}_i(0) \cdot \boldsymbol{v}_i(t) \rangle, \tag{8}$$

where $\boldsymbol{v}_i(t)$ is the velocity of atom $i$ at time $t$ and $\mathbf{W}_i$ is the virial tensor of atom $i$.

To obtain the frequency-dependent phonon mean free paths (MFPs), we further performed NEMD simulations with spectral analysis. In these simulations, two local thermostats maintained at different temperatures were applied as the heat source and heat sink, respectively, to establish a nonequilibrium steady state with a constant heat flux through the system. The resulting length-dependent spectral conductance was then used to extract the phonon MFP spectrum. Specifically, the thermal conductance between the two thermostats separated by a distance $L$ is given by[33, 35, 36]

$$G^{zz}(L) = \frac{J^{z}}{\Delta T V}, \tag{9}$$

where $\Delta T$ is the temperature difference between the heat source and heat sink. Then, the effective thermal conductivity for a system of finite transport length $L$ can be written as:

$$\kappa^{zz}(L) = G^{zz}(L)L = \frac{J^{z}}{(\Delta T/L)V}. \tag{10}$$

Similarly, the thermal conductance can be spectrally decomposed as:

$$G^{zz}(L) = \int_{0}^{\infty} \frac{d\omega}{2\pi} G^{zz}(L,\omega), \tag{11}$$

with the frequency-resolved thermal conductance expressed as:

$$G^{zz}(L,\omega) = \frac{2}{\Delta T V}\int_{-\infty}^{\infty} \mathrm{d}t\, \mathrm{e}^{i\omega t} K^{z}(t)\,. \tag{12}$$

In the ballistic limit, corresponding to $L \to 0$, the spectral ballistic conductance is defined as:

$$G_{\mathrm{b}}^{zz}(\omega) = G^{zz}(L \to 0, \omega). \tag{13}$$

The frequency-dependent phonon MFP can then be obtained from the ratio of the spectral thermal conductivity to the ballistic conductance as:

$$\lambda^{zz}(\omega) = \kappa^{zz}(\omega)/G_{\mathrm{b}}^{zz}(\omega). \tag{14}$$

Using the spectral MFP, the length-dependent spectral thermal conductivity can be evaluated as:

$$\kappa^{zz}(L,\omega) = \frac{\kappa^{zz}(\omega)}{1+\lambda^{zz}(\omega)/L}, \tag{15}$$

and the total length-dependent thermal conductivity is finally obtained by integrating over all phonon frequencies:

$$\kappa^{zz}(L) = \int_0^\infty \frac{\mathrm{d}\omega}{2\pi}\kappa^{zz}(L,\omega). \tag{16}$$

For the HNEMD calculations, the axial thermal conductivity of each $M_6X_6$ nanowire was calculated using the trained NEP and evaluated as the $zz$ component of the thermal conductivity tensor. All six $M_6X_6$ nanowire models were constructed with an axial length of approximately 300 nm, and a vacuum spacing of 100 Å was introduced in both the $x$ and $y$ directions to minimize artificial interactions between periodic images. During the NPT equilibration, only the axial cell dimension was allowed to relax at zero axial pressure, while the transverse cell dimensions were kept fixed. The corresponding size convergence is discussed in Section 3 of the Supporting Information. To report the HNEMD results in the conventional unit of $\mathrm{Wm^{-1}K^{-1}}$, an effective nanowire volume was defined as $V = AL_z$, where $L_z$ is the axial length and $A$ is the effective cross-sectional area. As illustrated in **Figure 1**, the cross section of each $M_6X_6$ nanowire was approximated as a circle, giving $A = \pi r^2$, where $r$ is the effective nanowire radius. The radii used for $Mo_6S_6$, $Mo_6Se_6$, $Mo_6Te_6$, $W_6S_6$, $W_6Se_6$, and $W_6Te_6$ were 2.87, 3.03, 3.25, 2.90, 3.07, and 3.20 Å, respectively.

Following the above definition of the effective volume, the HNEMD simulations were carried out using the trained NEP within the GPUMD[33] package according to the following protocol. Each system was initialized at 300 K and equilibrated for 1 ns in the NPT ensemble at 300 K and zero pressure, followed by an additional 1 ns equilibration in the NVT ensemble with a Nosé–Hoover chain thermostat. A time step of 1 fs was used throughout all simulations. The HNEMD production run was then conducted for 20 ns in the NVT ensemble, during which the external driving field was applied along the axial $z$ direction. The time evolution of the calculated thermal conductivity is shown in Section 4 of the Supporting Information, confirming that the HNEMD results reached a well-converged steady state. For each $M_6X_6$ nanowire, ten independent simulations with different initial velocity distributions were performed. The reported thermal conductivity was obtained by averaging the results from these independent trajectories, and the statistical uncertainty was evaluated from the variation among the individual simulations.

## 2.3. Vibrational DOS Calculations

The vibrational DOS was calculated from the Fourier transform of the velocity autocorrelation function (VACF). For each atomic species $s$, the vibrational DOS was evaluated as:[37]

$$g_s(\omega) = \frac{1}{N_s}\sum_{i\in s}\left(\int_{-\infty}^{+\infty}\langle \mathbf{v}_i(0)\cdot \mathbf{v}_i(\mathrm{t})\rangle\, e^{i\omega t}\mathrm{d}t\right), \tag{17}$$

where $N_s$ is the number of atoms of species $s$, $\mathbf{v}_i(t)$ is the velocity vector of atom $i$ at time $t$, and $\langle\cdot\rangle$ denotes the ensemble average. This formulation allows the total vibrational spectrum to be decomposed into contributions from the metal and chalcogen atoms.

For the vibrational DOS calculations, each $M_6X_6$ nanowire was simulated using the trained NEP within the GPUMD[33] package. Each system was first initialized at 300 K and equilibrated for 500 ps in the NPT ensemble at 300 K and zero pressure, followed by an additional 500 ps equilibration in the NVT ensemble. A time step of 1 fs was used throughout the simulations. The production trajectory was then generated from a 500 ps NVE simulation, during which the VACFs were accumulated for the subsequent vibrational DOS analysis.

## 2.4. Phonon SED Calculations

To characterize the phonon dispersion properties of the 1D materials, we computed the phonon spectral energy density (SED)[28] as a function of wave vector $\boldsymbol{k}$ and angular frequency $\omega$. In MD simulations, the SED is evaluated as:[38-40]

$$\Phi(\boldsymbol{k},\omega) = \frac{1}{4\pi\tau_0 N}\sum_{\alpha}^{\{x,y,z\}}\sum_{b}^{B} m_b \times \left|\int_0^{\tau_0}\sum_{n_{x,y,z}}^{N} v_\alpha\binom{n_{x,y,z}}{b};t\Big) \times \exp\left[i\boldsymbol{k}\cdot \boldsymbol{r}\binom{n_{x,y,z}}{0} - i\omega t\right]\mathrm{d}t\right|^2 . \tag{18}$$

Here, $\tau_0$ denotes the total simulation time, $N$ is the number of unit cells included in the sampled supercell, and $b$ labels the atoms within an individual unit cell. The quantity $m_b$ represents the mass of atom $b$, while $n_{x,y,z}$ identifies the unit cell along the $x$, $y$, and $z$ directions. The term $v_\alpha\binom{n_{x,y,z}}{b};t)$ is the $\alpha$-component of the velocity of atom $b$ in unit cell $n$ at time $t$, and $\boldsymbol{r}\binom{n_{x,y,z}}{0}$ denotes its equilibrium position.

For the SED calculations, the equilibrium structures of the $M_6X_6$ nanowires were expanded into $1 \times 1 \times 200$ supercells along the axial direction to provide sufficient resolution in the 1D reciprocal space. A vacuum spacing of 100 Å was introduced in both the $x$ and $y$ directions to suppress artificial interactions between periodically repeated nanowires. Each system was first equilibrated for 200 ps in the NPT ensemble at 300 K with the axial pressure maintained at zero, followed by a further 500 ps equilibration in the NVT ensemble. A time step of 1 fs was used throughout the simulations. Subsequently, a 500 ps production run was performed in the NVE ensemble, during which the atomic velocities and positions were collected every 8 fs for the subsequent spectral energy density analysis. The spectral energy density spectra and the corresponding Lorentzian fitting were obtained using the PYSED package.[28]

## 3. RESULTS AND DISCUSSION

### 3.1. Machine-learned Interatomic Potential

A reliable NEP model depends critically on the quality and diversity of the training dataset. As illustrated in **Figure 1**, the $M_6X_6$ nanowires considered in this work share the same 1D structural framework but involve different transition-metal and chalcogen compositions. Accordingly, the training dataset was constructed to include representative configurations of all six $M_6X_6$ nanowires, including $Mo_6S_6$, $Mo_6Se_6$, $Mo_6Te_6$, $W_6S_6$, $W_6Se_6$, and $W_6Te_6$, thereby ensuring adequate coverage of the relevant configurational and chemical spaces. The reference structures include both near-equilibrium and distorted configurations generated from random perturbations and DFT-MD sampling. Further details of the dataset generation and composition are provided in the Methods section. Single-point DFT calculations were performed for all reference configurations to obtain the total energies, atomic forces, and virial tensors used for NEP training. As shown in **Figure 3**, the resulting NEP model reproduces the DFT reference data with high fidelity. The RMSEs for the training set are 1.48 meV/atom for energy, 104.37 meV/Å for force, and 28.36 meV/atom for virial, while those for the test set are 1.48 meV/atom, 104.95 meV/Å, and 27.97 meV/atom, respectively.

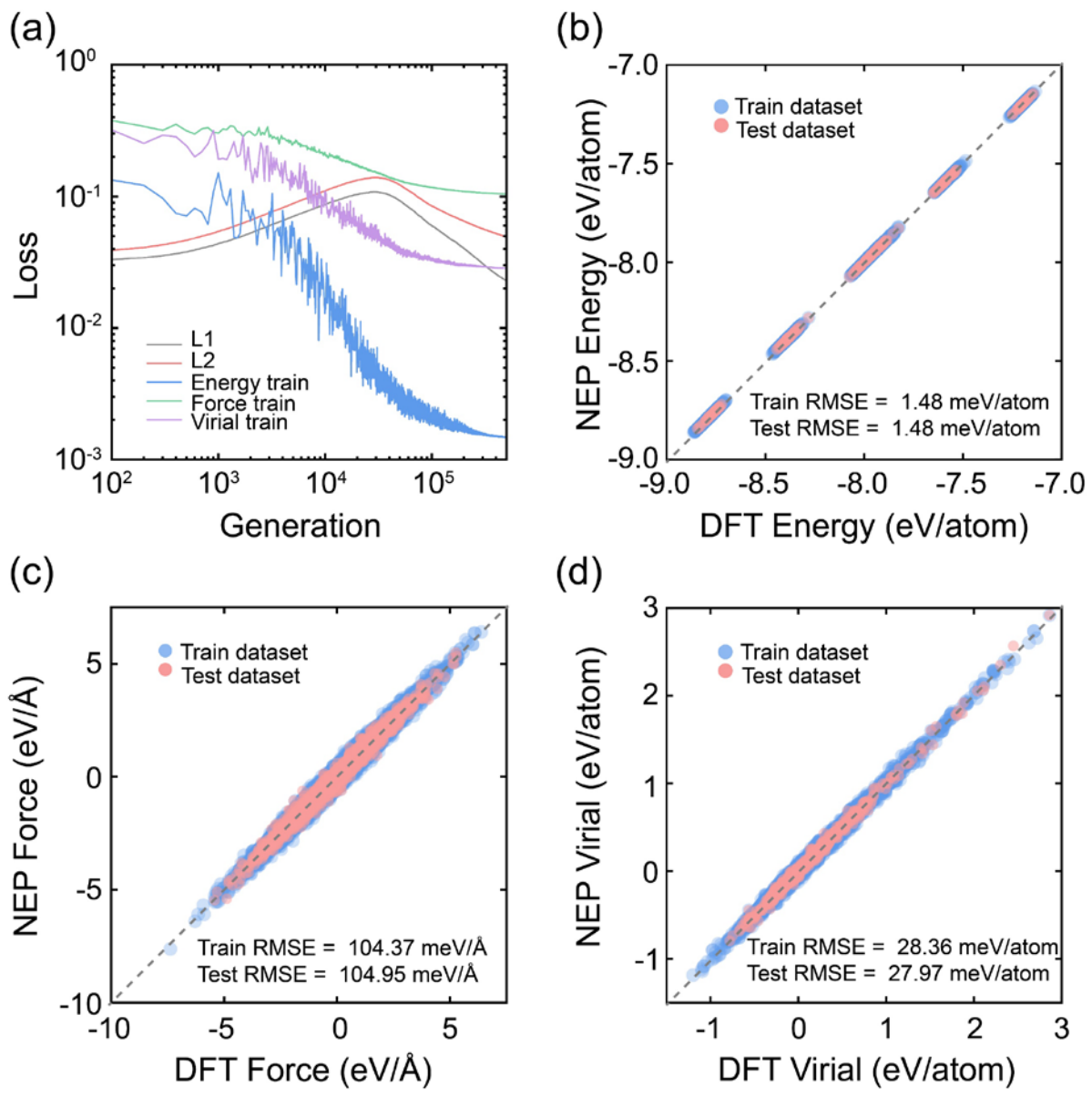

**Figure 3.** NEP model training and validation. (a) Evolution of the RMSEs of energy, force, and virial during training as a function of training generation. Comparison of (b) energy, (c) force, and (d) virial predicted by NEP against DFT reference data for the initial training dataset.

## 3.2. Structural Validation of the NEP Model

To further validate the structural reliability of the trained NEP model, we compared the optimized geometrical parameters predicted by NEP with previously reported DFT results. In particular, the axial lattice parameter $a$ and the nanowire height $h$, as defined in **Figure 1**b and c, were selected as representative structural descriptors. These two parameters characterize the periodicity along the 1D direction and the transverse size of the $M_6X_6$ nanowires, respectively. As summarized in **Table 1**, the predicted values by NEP of $a$ and $h$ agree well with the reference data for all six $M_6X_6$ nanowires.

The local bonding characteristics were further examined by comparing the partial radial distribution functions (RDFs) obtained from NEP and PBE calculations. Specifically, the Mo-S, Mo-Se, Mo-Te, W-S, W-Se, and W-Te RDFs were evaluated for the corresponding nanowire systems. As shown in **Figure 4**, the NEP results closely reproduce the PBE reference data.

These results demonstrate that the present NEP model can accurately describe not only the equilibrium geometries but also the characteristic bond lengths and local coordination environments of the $M_6X_6$ nanowires.

**Table 1** Optimized axial lattice parameter $a$ (Å) and the nanowire height $h$ (Å), as marked in **Figure 1**b and c, for six $M_6X_6$ nanowires. The structural parameters calculated using DFT and NEP are compared with available experimental values. The DFT data were taken from ref. 41. The experimental data for $Mo_6S_6$, $Mo_6Se_6$ and $W_6S_6$ were obtained from ref. 42, whereas the experimental data for $Mo_6Te_6$ was taken from ref. 21.

| Nanowire systems | $a$ (Å) | | | $h$ (Å) | |
|---|---|---|---|---|---|
| | DFT | Exp. | NEP | DFT | NEP |
| $Mo_6S_6$ | 4.35 | 4.5 | 4.31 | 4.30 | 4.31 |
| $Mo_6Se_6$ | 4.44 | 4.5 | 4.38 | 4.53 | 4.54 |
| $Mo_6Te_6$ | 4.58 | 4.8 | 4.58 | 4.87 | 4.87 |
| $W_6S_6$ | 4.37 | 4.4 | 4.34 | 4.35 | 4.36 |
| $W_6Se_6$ | 4.46 | — | 4.41 | 4.59 | 4.60 |
| $W_6Te_6$ | 4.56 | — | 4.53 | 4.95 | 4.92 |

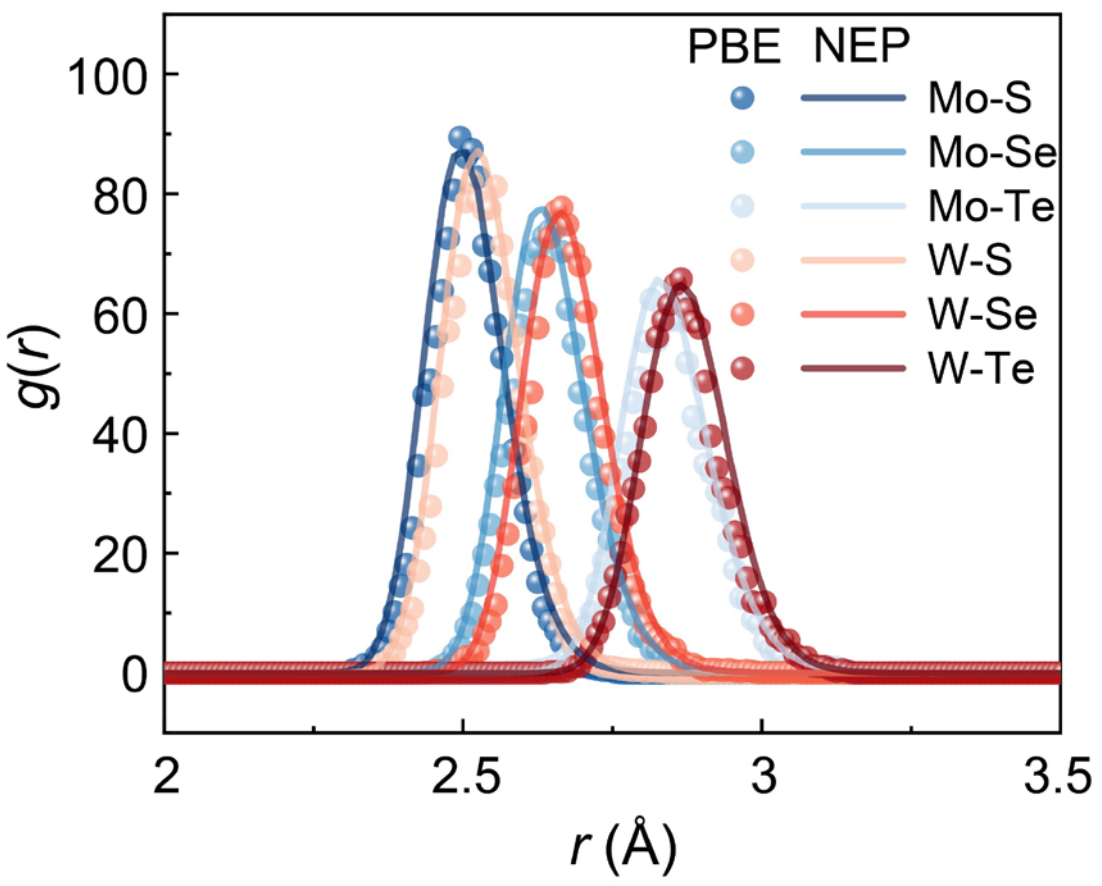

**Figure 4.** Comparison of the partial radial distribution functions $g(r)$ for the metal and chalcogen atom pairs in the six $M_6X_6$ nanowires, as obtained from PBE and NEP calculations. The Mo-S, Mo-Se, Mo-Te, W-S, W-Se, and W-Te RDFs correspond to the respective nanowire systems. Symbols denote the PBE results, while solid lines represent the NEP predictions

### 3.3. Thermal Transport of $M_6X_6$ Nanowires

Following the development of the NEP model, we calculated the intrinsic axial thermal conductivity, $\kappa$, of the six $M_6X_6$ nanowires at 300 K using the HNEMD method implemented in GPUMD. The external driving field was applied along the nanowire axis. As shown in **Figure 5**, the predicted thermal conductivities exhibit a pronounced dependence on elemental composition. The Mo-based and W-based nanowires display distinct composition-dependent trends. For the $Mo_6X_6$ series, $\kappa$ increases from 294 ± 14 $Wm^{-1}K^{-1}$ for $Mo_6S_6$ to 357 ± 12 $Wm^{-1}K^{-1}$ for $Mo_6Se_6$, and then decreases to 235 ± 15 $Wm^{-1}K^{-1}$ for $Mo_6Te_6$. In contrast, th$W_6X_6$ series shows a monotonic decrease in $\kappa$, from 263 ± 11 $Wm^{-1}K^{-1}$ for $W_6S_6$ to 242 ± 14 $Wm^{-1}K^{-1}$ for $W_6Se_6$ and further to 192 ± 12 $Wm^{-1}K^{-1}$ for $W_6Te_6$. For the same chalcogen element, replacing Mo with W consistently reduces the axial thermal conductivity, with the most pronounced reduction observed for the selenide nanowires.

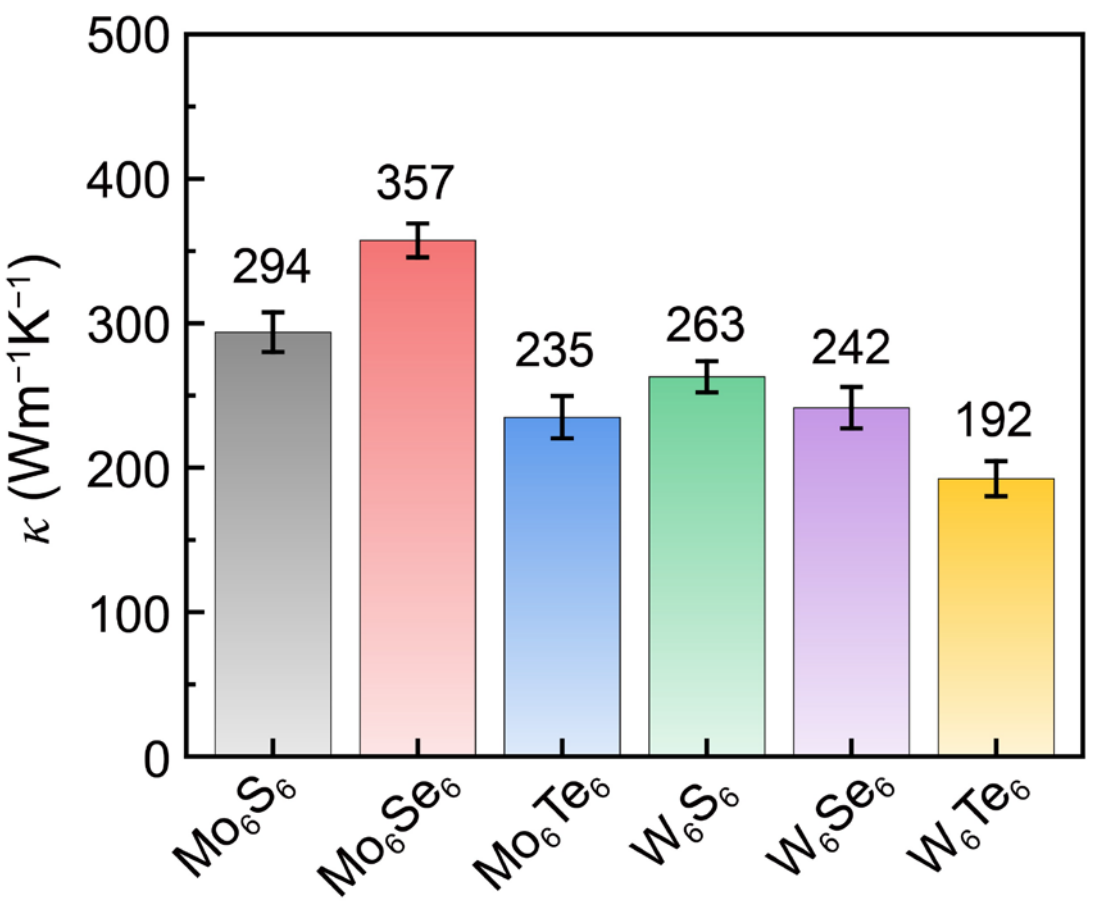

**Figure 5.** Axial thermal conductivity of $M_6X_6$ nanowires. Thermal conductivity $\kappa$ of six $M_6X_6$ nanowires calculated by HNEMD simulations at 300 K, including $Mo_6S_6$, $Mo_6Se_6$, $Mo_6Te_6$, $W_6S_6$, $W_6Se_6$, and $W_6Te_6$. The bars with different colors represent different nanowire compositions, and the error bars denote the statistical uncertainties estimated from ten independent simulations.

To identify the frequency ranges responsible for the axial thermal conductivity, we further performed spectral decomposition of the heat transport. As shown in **Figure 6**a, the frequency-resolved thermal conductivity $\kappa(\omega)$ is mainly concentrated in the low-frequency region for all six $M_6X_6$ nanowires, indicating that axial heat conduction is dominated by low-frequency phonons. The cumulative thermal conductivity shown in the inset further confirms that most of the total $\kappa$ is accumulated within the low-frequency phonon spectrum. Among the six systems, $Mo_6Se_6$ exhibits the largest low-frequency spectral contribution, consistent with its highest axial thermal conductivity, whereas $W_6Te_6$ shows the weakest contribution and the lowest $\kappa$. In addition, the frequency-dependent phonon mean-free path $\lambda(\omega)$ in **Figure 6**b decreases rapidly with increasing frequency, suggesting that low-frequency phonons possess longer effective transport lengths and therefore dominate the length-dependent thermal response. The inset of **Figure 6**b shows that $\kappa(L)$ gradually increases with transport length and approaches the intrinsic value at sufficiently long lengths.

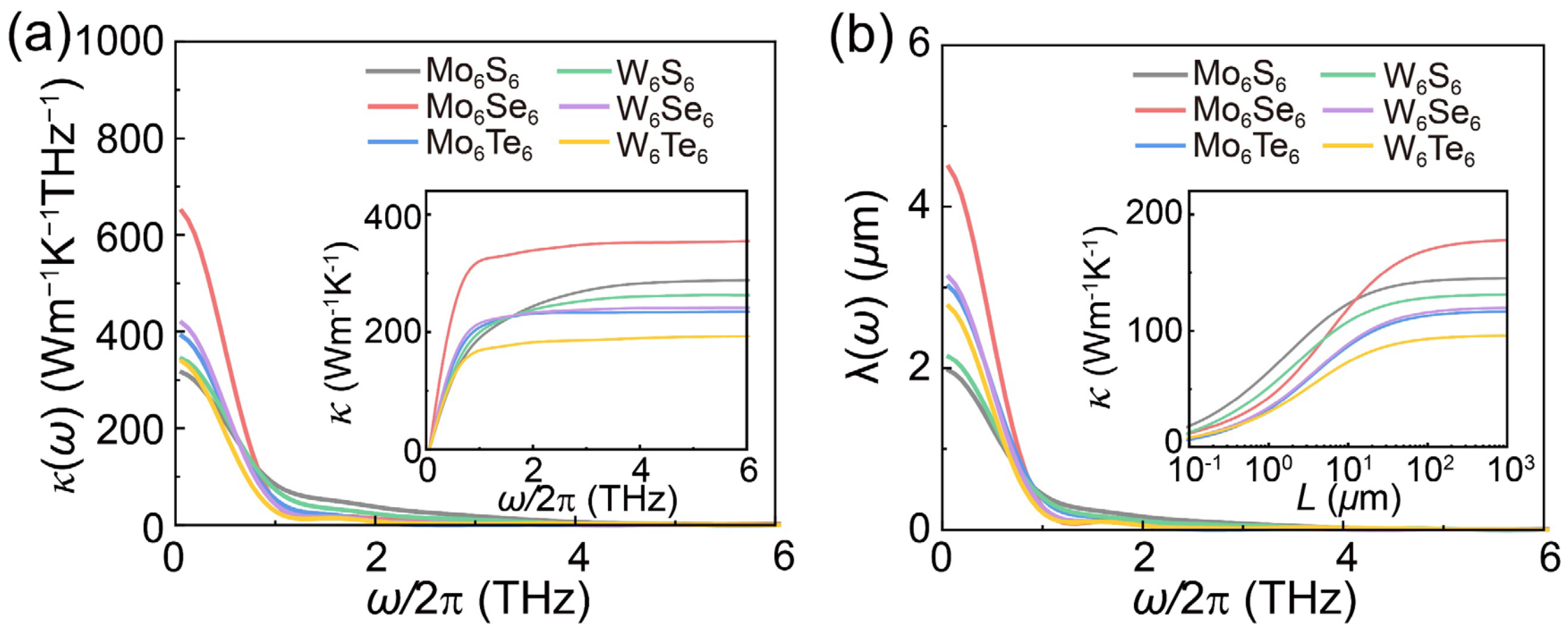

**Figure 6.** Spectral thermal transport properties of $M_6X_6$ nanowires. (a) Frequency-resolved axial thermal conductivity $\kappa(\omega)$ of the six $M_6X_6$ nanowires obtained from HNEMD simulations at 300 K. The inset shows the cumulative axial thermal conductivity as a function of phonon frequency. (b) Frequency-dependent phonon mean-free path $\lambda(\omega)$. The inset shows the corresponding length-dependent axial thermal conductivity $\kappa(L)$. The gray, red, blue, green, purple, and orange curves represent $Mo_6S_6$, $Mo_6Se_6$, $Mo_6Te_6$, $W_6S_6$, $W_6Se_6$, and $W_6Te_6$, respectively.

The above spectral results indicate that the difference in thermal conductivity among $M_6X_6$ nanowires is primarily associated with low-frequency phonon transport channels. However, the spectral decomposition alone does not directly reveal how elemental composition modifies the underlying phonon modes. Therefore, as shown in **Figure 7**, we first examined the vibrational DOS projected onto individual atomic species to evaluate the vibrational coupling between the metal and chalcogen atoms.[43] The spectral overlap between the metal- and chalcogen-projected DOS provides a qualitative measure of vibrational matching, which is expected to facilitate phonon transport. For the Mo-based nanowires, the integrated DOS-overlap area below 2 THz reaches its maximum in $Mo_6Se_6$. When normalized to $Mo_6S_6$, the overlap areas are 1.00, 2.04, and 1.30 for $Mo_6S_6$, $Mo_6Se_6$, and $Mo_6Te_6$, respectively. The enhanced overlap in $Mo_6Se_6$ indicates stronger Mo—Se vibrational matching, consistent with its highest axial thermal conductivity. However, this descriptor fails to capture the lower thermal conductivity of $Mo_6Te_6$ relative to $Mo_6S_6$, indicating that DOS overlap alone can only partially explain the composition-dependent thermal transport trend. Moreover, this simple overlap-based interpretation does not fully account for the W-based series, where the DOS overlap does not directly follow the monotonic decrease in thermal conductivity from $W_6S_6$ to $W_6Te_6$. This discrepancy suggests that vibrational matching alone is insufficient to determine the composition-dependent thermal transport.

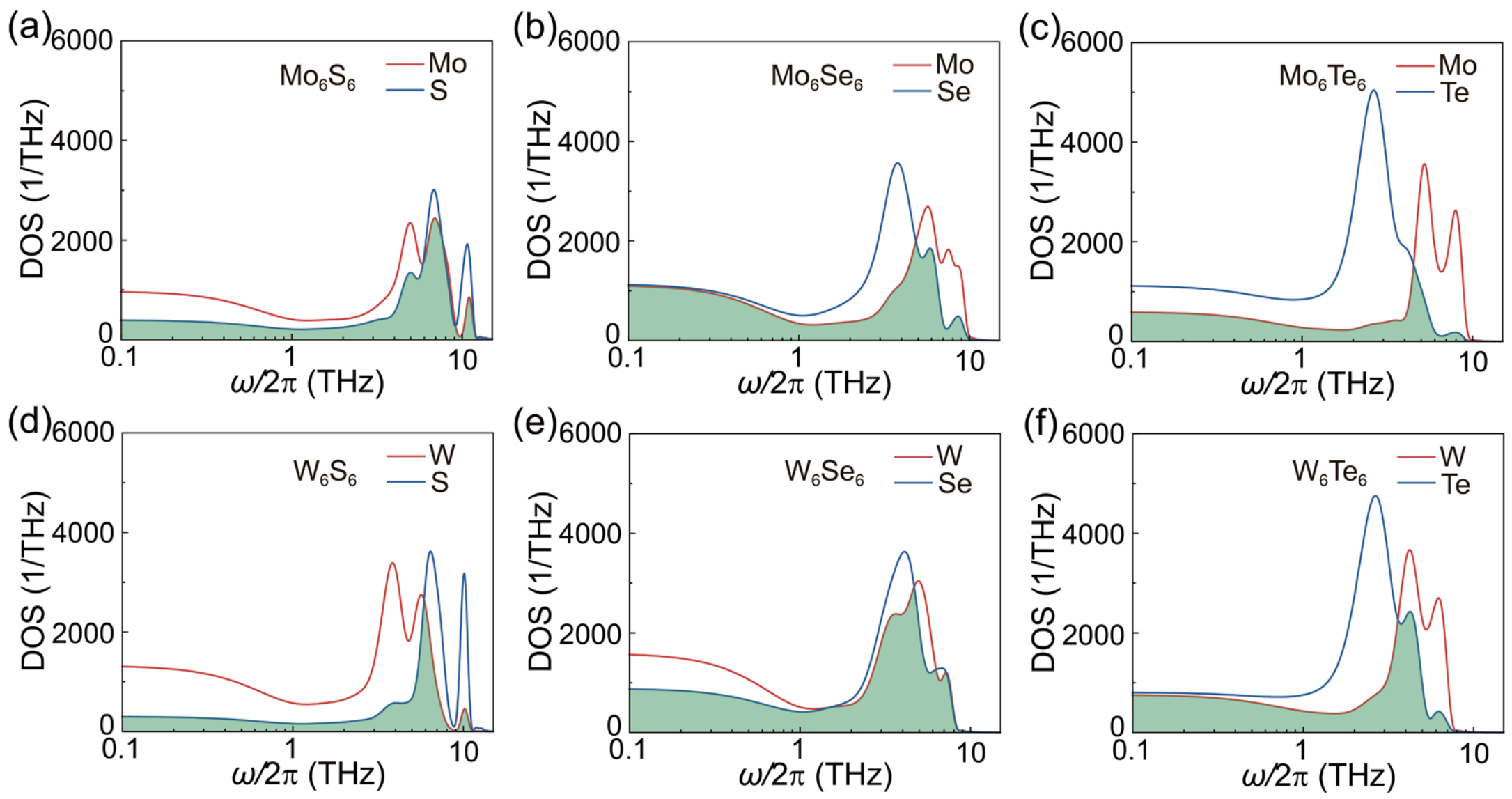

**Figure 7.** Vibrational DOS projected onto individual atomic species for $M_6X_6$ nanowires. Vibrational DOS of (a) $Mo_6S_6$, (b) $Mo_6Se_6$, (c) $Mo_6Te_6$, (d) $W_6S_6$, (e) $W_6Se_6$, and (f) $W_6Te_6$ nanowires, calculated from the Fourier transform of the velocity autocorrelation function. The red and blue curves denote the metal atom and chalcogen atom contributions, respectively. The shaded regions indicate the spectral overlap between the metal and chalcogen vibrational contributions.

To further clarify this competition, we analyzed the phonon spectra using SED calculations. As shown in **Figure 8**, replacing S with heavier Se and Te atoms induces systematic phonon softening in both Mo- and W-based nanowires, as evidenced by the downward shift of the phonon branches. This trend is consistent with the mass dependence of lattice vibrations, where heavier atoms generally reduce phonon frequencies when the bond stiffness does not increase correspondingly.[44, 45] Although phonon softening can enhance the population of low-frequency vibrational modes, it also tends to reduce phonon group velocities (see **Figure 9**) and weaken the heat-carrying capability of dispersive acoustic branches [46]. This interpretation is also consistent with Slack's criterion, which suggests that high thermal conductivity is generally favored in simple crystals composed of light atoms and stiff chemical bonds [47]. Therefore, the thermal conductivity of $M_6X_6$ nanowires is governed by the balance between metal-chalcogen vibrational matching and the phonon softening caused by heavier chalcogen atoms.

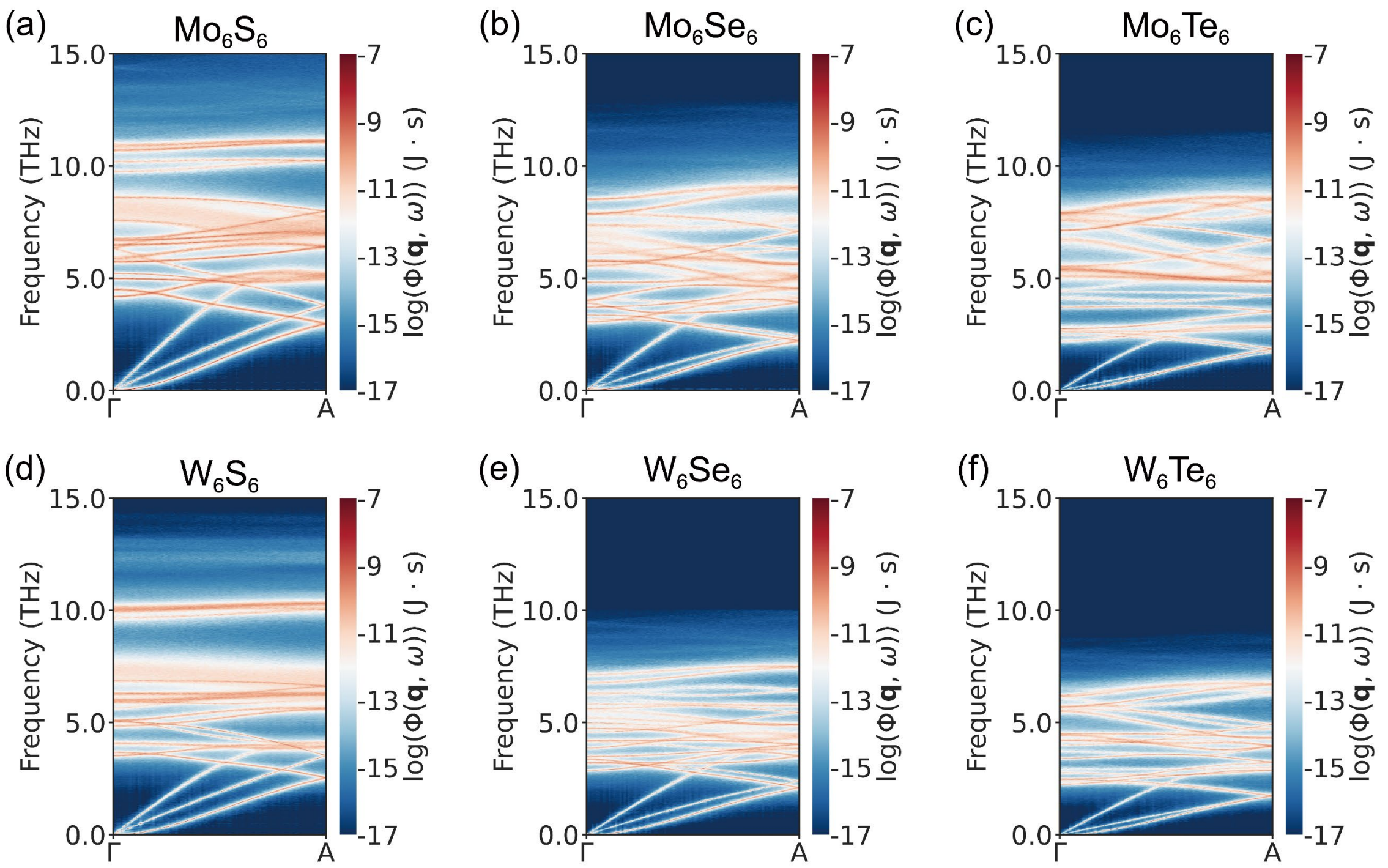

**Figure 8.** Spectral energy density analysis of six $M_6X_6$ nanowires. SED maps of (a) $Mo_6S_6$, (b) $Mo_6Se_6$, (c) $Mo_6Te_6$, (d) $W_6S_6$, (e) $W_6Se_6$, and (f) $W_6Te_6$ nanowires calculated along the axial Γ–A direction.

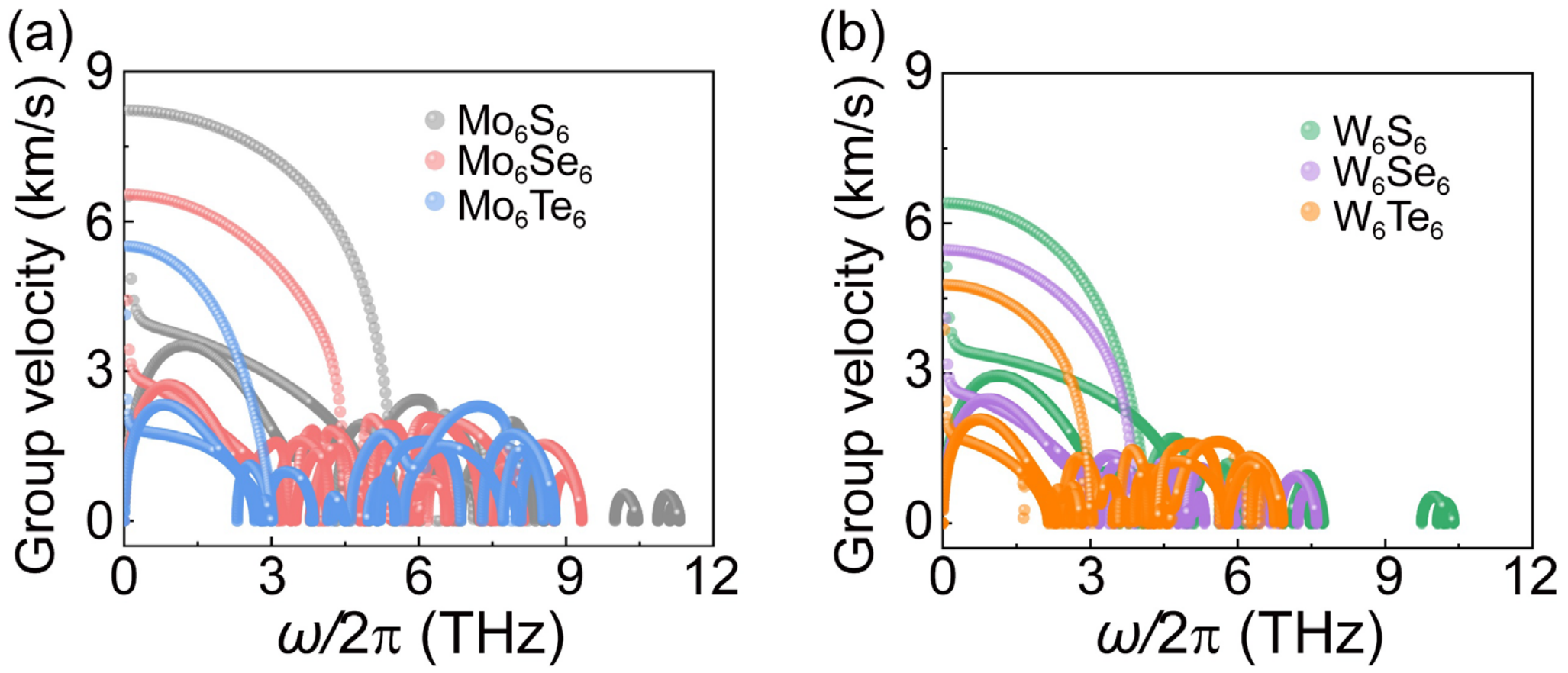

**Figure 9.** Phonon group velocity of of (a) $Mo_6X_6$ and (b) $W_6X_6$ nanowires, where X = S, Se, and Te. The harmonic force constants were obtained using finite-displacement calculations within the NEP-kappa workflow [48], with atomic forces predicted by the corresponding NEP models, while the phonon properties were evaluated using phono3py [49].

For the $Mo_6X_6$ series, $Mo_6Se_6$ achieves an optimal balance between these two effects. Compared with $Mo_6S_6$, Se substitution enhances the overlap between Mo- and chalcogen-projected vibrational spectra, while the phonon softening remains moderate, allowing efficient low-frequency heat-carrying modes to be preserved. This balance leads to the highest thermal conductivity in $Mo_6Se_6$. In contrast, Te substitution causes stronger phonon softening, reducing the effective transport capability of low-frequency branches and thus lowering the thermal conductivity of $Mo_6Te_6$.

For the $W_6X_6$ series, the larger atomic mass of W already results in a softer phonon spectrum than in the Mo-based counterparts. Further substitution of S by Se or Te therefore causes more pronounced softening. Consequently, although $W_6Se_6$ may exhibit favorable vibrational matching between W and Se atoms, the downward shift in phonon frequencies and the reduced contribution from dispersive phonon branches play a more important role. As a result, the thermal conductivity decreases monotonically from $W_6S_6$ to $W_6Te_6$. These results demonstrate that the distinct composition-dependent thermal transport trends in $Mo_6X_6$ and $W_6X_6$ originate from the balance between vibrational matching and the reduction of phonon frequencies associated with increasing atomic mass.

As discussed above, the SED and PDOS analyses indicate that chalcogen substitution simultaneously modifies the vibrational compatibility between metal and chalcogen atoms and shifts the phonon spectrum toward lower frequencies due to the increasing atomic mass. These two effects jointly determine the composition dependence of thermal conductivity. To further provide a compact phenomenological rationalization of the above interpretation, we constructed a simple phenomenological model to describe the competing effects of vibrational matching and phonon softening on the thermal transport of $M_6X_6$ nanowires. The vibrational matching effect is quantified using a dimensionless parameter that describes the relative mass difference between the metal and chalcogen atoms, as $\delta_{\mathrm{MX}} = |m_{\mathrm{M}} - m_{\mathrm{X}}|/(m_{\mathrm{M}} + m_{\mathrm{X}})$, where $m_{\mathrm{M}}$ and $m_{\mathrm{X}}$ denote the atomic masses of the metal and chalcogen atoms, respectively. A smaller $\delta_{\mathrm{MX}}$ indicates a smaller mass mismatch between the two atomic sublattices, which is expected to facilitate vibrational coupling and promote phonon transport. Meanwhile, the phonon softening induced by chalcogen substitution is described by a dimensionless mass-softening descriptor, $x_{\mathrm{MX}} = s_{\mathrm{M}} \frac{m_{\mathrm{MX}} - m_{\mathrm{MS}}}{m_{\mathrm{MS}}}$, with $s_{\mathrm{M}} = \frac{m_{\mathrm{M}}}{m_{\mathrm{Mo}}}$ and $m_{\mathrm{MX}} = m_{\mathrm{M}} + m_{\mathrm{X}}$. Here, $m_{\mathrm{MS}}$ denotes the combined mass of the S-based compound in each series. The factor $s_{\mathrm{M}}$ accounts for the stronger softening effect in nanowires containing W. By incorporating these two effects, the normalized thermal conductivity is expressed as:

$$\tilde{\kappa} = \exp\left[-a\left(x_{\mathrm{MX}} + \frac{1}{2}x_{\mathrm{MX}}^2\right)\right]\left(\frac{\delta_{\mathrm{MS}}}{\delta_{\mathrm{MX}}}\right)^b, \tag{19}$$

where $\tilde{\kappa} = \kappa_{\mathrm{MX}}/\kappa_{\mathrm{MS}}$, and the parameters $a$ and $b$ are two global phenomenological parameters. The parameter $a$ measures the sensitivity of thermal conductivity to phonon softening induced by the increased atomic mass, whereas $b$ measures the sensitivity to vibrational matching represented by the mass mismatch ratio. Simultaneous fitting of the four non-reference data points gives $a = 0.55$ and $b = 0.26$.

As shown in **Figure 10**, the phenomenological model reproduces the main trends obtained from HNEMD simulations. For the $Mo_6X_6$ series, the normalized thermal conductivity first increases from $Mo_6S_6$ to $Mo_6Se_6$ and then decreases for $Mo_6Te_6$. This indicates that Se substitution improves vibrational matching while maintaining moderate phonon softening, whereas Te substitution introduces stronger phonon softening that suppresses heat transport. In contrast, the $W_6X_6$ series shows a monotonic decrease from $W_6S_6$ to $W_6Te_6$. This behavior suggests that, because of the already large mass of W, further chalcogen substitution mainly enhances phonon softening, and the improvement in vibrational matching is insufficient to compensate for the reduced phonon frequencies. These results confirm that the different composition-dependent thermal transport trends in $Mo_6X_6$ and $W_6X_6$ nanowires can be consistently understood from the balance between vibrational matching and the reduction of phonon frequencies associated with increasing atomic mass.

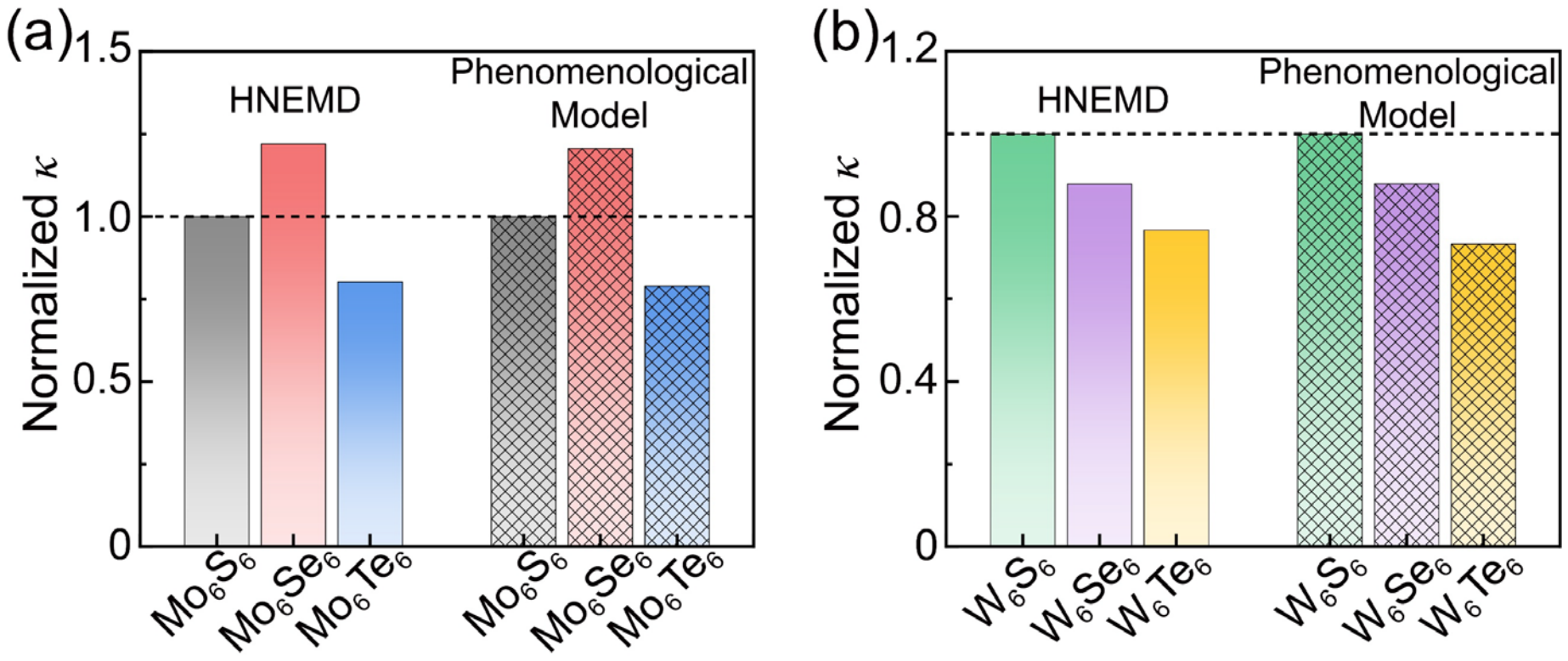

**Figure 10.** Comparison between HNEMD results and the phenomenological model. Normalized thermal conductivity of (a) $Mo_6X_6$ and (b) $W_6X_6$ nanowires obtained from HNEMD simulations and the phenomenological model. For each series, the thermal conductivity is normalized by the corresponding S-based compound, namely $Mo_6S_6$ for the $Mo_6X_6$ series and $W_6S_6$ for the $W_6X_6$ series.

## 4. CONCLUSION

In summary, we developed a multicomponent NEP model for six $M_6X_6$ nanowires (M = Mo or W; X = S, Se, or Te) and used HNEMD simulations to investigate their axial thermal transport. The trained model accurately reproduces DFT reference data and structural characteristics. At 300 K, the thermal conductivity exhibits distinct composition-dependent trends: $Mo_6Se_6$ shows the highest value of 357 ± 12 $Wm^{-1}K^{-1}$, while $W_6Te_6$ has the lowest value of 192 ± 12 $Wm^{-1}K^{-1}$. The $Mo_6X_6$ series displays a nonmonotonic variation, whereas the thermal conductivity of $W_6X_6$ series decreases monotonically with increasing chalcogen mass. Spectral analyses show that heat transport is dominated by low-frequency phonons and governed by the competition between metal-chalcogen vibrational matching and phonon softening caused by heavier chalcogen atoms. A phenomenological model incorporating both effects captures the main HNEMD trends, offering a unified physical interpretation of the composition-dependent thermal transport in $M_6X_6$ nanowires. These results clarify the relationship among elemental composition, phonon characteristics and thermal conductivity, and provide an atomistic basis for designing nanowire bundles and arrays with tunable thermal transport properties.

## Supporting Information

Supporting Information contains the following sections: Construction of the training dataset for NEP model, convergence test of the driving force parameter in HNEMD simulations, axial size convergence test in HNEMD simulations, and time convergence of axial thermal conductivity in HNEMD simulations.

### Author Contributions

W.J. conceived the research and supervised the project. W.J. and H.B. performed all MD simulations. All authors contributed to data analysis and manuscript writing.

### Notes

The authors declare no competing financial interest.

### Acknowledgments

W.J. acknowledges the National Key Research and Development Program of China (Grant No. 2022YFA1203602) and the Fundamental Research Funds for the Central Universities. Computations were conducted at the National Supercomputer TianHe-1(A) Center in Tianjin.

### Data availability

All input and output files related to the training of the transition metal chalcogenide $M_6X_6$ nanowires (M = Mo or W; X = S, Se, or Te) are freely available via the Zenodo link: https://doi.org/10.5281/zenodo.22171223.